\documentclass[twocolumn, 10pt, letterpaper]{article}
\usepackage[pass, margin=1in]{geometry} 
\usepackage{ibpsa} 
\usepackage{pslatex}
\usepackage{achicago}
\usepackage{amsmath, amssymb}
\usepackage{graphicx}
\usepackage{fancyhdr}
\usepackage{ifthen}
\usepackage[dvipsnames]{xcolor}
\usepackage{makecell}
\usepackage{graphicx,subcaption}

\renewcommand{\eqref}[1]{(\ref{#1})}

\begin{document}
\date{}

\ifthenelse{\value{page}=1}{\setlength{\textheight}{8.5in}}{\setlength{\textheight}{9in}}

\title{\vspace{-0.4in} \titlefont 
    Sustainability through Optimal Design of Buildings for \\ Natural Ventilation using Updated Comfort and Occupancy Models \vspace{-0.2in} 
}

\author{%
    \authorfont{~}\\
    \authorfont{Jihoon Chung$^1$, Nastaran Shahmansouri$^2$, Rhys Goldstein$^2$, James Stoddart$^3$, and John Locke$^3$}\\
    \authorfont{$^1$Rensselaer Polytechnic Institute, New York, USA}\\
    \authorfont{$^2$Autodesk Research, Toronto, Canada}\\
    \authorfont{$^3$Autodesk Research, New York, USA}\\
    \authorfont{~}\\ 
    \vspace{-0.55in} 
}

\maketitle
\thispagestyle{empty}

\section{Abstract}
This paper explores the benefits of incorporating natural ventilation (NV) simulation into a generative process of designing residential buildings to improve energy efficiency and indoor thermal comfort. Our proposed workflow uses the Wave Function Collapse algorithm to generate a diverse set of plausible floor plans. It also includes post-COVID occupant presence models while incorporating adaptive comfort models. We conduct four sets of experiments using the workflow, and the simulated results suggest that multi-mode cooling strategies combining conventional air conditioning with NV can often significantly reduce energy use while introducing only slight reductions in thermal comfort. 

\section{Introduction}
Buildings account for roughly 30\% of global final energy use and over 55\% of global electricity consumption \cite{iea2019critical,iea2022final}. Between 1990 and 2016, the energy consumption for space cooling in buildings more than tripled, and the cooling load is the primary contributor to rising energy demand in the building sector. Furthermore, about 70\% of the air conditioning (AC) units sold annually are for residential buildings, and their output capacity accounts for 70.3\% (848 GW) of the total output capacity of AC units across building types \cite{iea2018future}. Strategies for reducing cooling loads in residential buildings have the potential to significantly mitigate energy consumption and carbon dioxide emissions in the building sector.

Passive techniques are some of the oldest ways to enhance occupants' thermal comfort and decrease energy demand from heating and cooling loads \cite{tejero2016assessing,cheng2018evaluating}. Despite their long-standing history, these techniques remain promising due to their low installation and operating expenses, minimal carbon dioxide emissions, high durability, and low noise levels, in comparison to mechanical systems \cite{prajongsan2012enhancing,zhang2021critical}. 

Natural ventilation (NV) is among the most potentially beneficial passive cooling techniques available for improving energy efficiency, indoor thermal comfort, and air quality in residential buildings \cite{mushtaha2021impact,fereidani2021review,laurini2018passive,wen2023framework}. NV systems consist of purpose-built openings for ventilation, such as wind-induced airflow, the buoyancy (stack) effect, or hybrid ventilation, to provide convective heat dissipation and increase evaporative heat loss from human bodies \cite{prajongsan2012enhancing,zhang2021critical}. NV has been shown to significantly decrease building energy use in many countries with varying climate features. For instance, in Spain, combining the air-conditioning system with NV resulted in a reduction of up to 29.5\% of HVAC operating hours \cite{castano2018adaptive}. In Andalusian schools, NV systems reduced annual primary energy usage by 18\% to 33\% \cite{gil2017natural}. Additionally, NV contributed to a 24\% reduction in energy use of air conditioning systems in Hong Kong \cite{gil2017natural,yik2010energy}. NV is therefore seen as a key passive cooling technique that can help minimize energy consumption compared with conventional AC systems.

Designing an effective NV system remains difficult due to its dependence on the climate and various design variables, such as outdoor temperature, wind direction, building shape, orientation, plan layout, and window size \cite{wen2023framework,cheng2018evaluating,zhou2014design}. Evaluating NV performance can be challenging due to its complexity, making it difficult for architects and stakeholders to make design decisions and adopt NV systems. Moreover, buildings designed and constructed for conventional cooling can be challenging to later modify to support NV systems \cite{tejero2016assessing}. For this reason, integrating NV simulation with an automated generative geometry modeler may be helpful as a mean of exploring diverse building forms with high NV performance during the early design stage \cite{wen2023framework,chen2016holistic,malkawi2016predicting,li2021design}.

This paper aims to advance the modeling of NV solutions by exploring a simulation-based workflow with the following features: 1) the utilization of the Wave Function Collapse (WFC) algorithm to generate mid-rise residential buildings with a diverse set of floor plan layouts; 2) the incorporation of up-to-date occupant presence models that reflect post-COVID behavior in residential buildings; and 3) the adoption of thermal comfort models that account for the thermal sensations of residential building occupants as reported in the literature. We envision the presented workflow as a step toward creating a generative tool for planning and optimizing naturally ventilated buildings during the early design stage. As a preliminary step toward this goal, we conduct four sets of simulation-based experiments using the workflow across various climate zones to assess the impact of different building layouts, orientations, and cooling strategies on energy consumption and occupant comfort in buildings. The experiments provide insight into the use of WFC-based geometry generation when comparing a) conventional AC strategies, b) NV strategies, and c) multi-mode (MM) strategies that combine AC and NV.

\section{Related Work}
Recently, simulation tools have been combined with generative geometry modeling tools to explore optimal combinations of multiple design parameters \cite{grygierek2018multi,weerasuriya2019holistic,omrani2017natural}. This process involves iterative evaluation and feedback on the impact of NV on building performance, with the goal of finding an optimal building design. For example, Zhou et al. (2014) explored optimal building orientation and spacing to reduce the age of indoor air, while Grygierek and Ferdyn-Grygierek (2018) investigated optimal combinations of building orientation, glazing type, window area, and thermal insulation \cite{zhou2014design,grygierek2018multi}. Yoon et al. (2020) focused on exploring window-opening positions to optimize NV performance, while Ameur et al. (2020) explored window-opening areas \cite{yoon2020optimization,ameur2020optimization}. Yin et al. (2023) optimized window type, window position, opening fraction, and furniture placement to maximize the air change rate in a room \cite{yin2023multi}. The recent NV-related generative modeling research mainly focused on building characteristics, such as window dimension and position, or building orientation; however, only few studies considered occupant-related factors to precisely evaluate NV, including occupant comfort models and occupancy schedules.

To reflect precise occupants' thermal comfort and occupancy schedules, several comfort and occupancy models have been proposed. One of the standardized comfort models is the Predicted Mean Vote (PMV), which predicts the mean value of a large group of people by considering indoor temperature, thermal radiation, relative humidity, air velocity, occupants' metabolic rate, and clothing insulation \cite{international2005ergonomics}. However, this model estimates occupants' comfort based on the heat balance model, ignoring the psychological dimension of adaptation, increasing gap between the estimated and actual comfort temperature in naturally ventilated buildings where occupants control operable windows \cite{de2002thermal}. To overcome this limitation, ASHRAE RP-884 collected 21,000 sets of raw field data, and used statistic analysis to develop regression models for "adaptive comfort". These models showed higher accuracy for predicting thermal comfort in naturally ventilated buildings than the PMV model; however, they may have limitations when applied to naturally ventilated or multi-mode residential buildings due to the relatively small number of such building in the dataset \cite{de2020review}. Moreover, an extensive review conducted by \citeN{kim2021mixed} concluded that the adaptive comfort model consistently outperformed the PMV model with a better alignment between the occupants' actual comfort votes in multi-mode buildings and predicted comfort values although occupants in the multi-mode buildings have reported higher satisfaction levels and wider comfort ranges than existing standardized models.

Since the outbreak of COVID-19, traditional workplace practices and employees' workplace preferences have undergone a drastic transformation. In the United States, for example, 48\% of Americans still anticipate having the option to work from home after the pandemic, which represents approximately a 30\% increase compared to the pre-COVID period  \cite{yang2022effects,javadinasr2022long}. In the case of Canada, 32\% of Canadian employees worked most of their hours from home at the beginning of 2021, while only 4\% did in 2016 \cite{mehdi2021working}. In the Netherlands, 27\% of teleworkers responded in April 2020 that they expected to work more from home in the future \cite{de2020covid}. The growing number of teleworkers has also altered the pattern of HVAC usage and electricity consumption in residential buildings \cite{sirati2023household,villeneuve2021new}. Currently, occupancy profiles from the 1989 version of ASHRAE 90.1 and ASHRAE Advanced Energy Design Guide are among the most commonly used for building energy modeling software programs \cite{mitra2021variation}. However, these profiles appear outdated in light of post-COVID occupancy patterns reported in the recent literature. We therefore strive to use up-to-date occupancy models, and we submit that this should become standard practice to ensure that the COVID pandemic's lasting effects on daily life are reflected in building energy and comfort simulations.

In summary, despite the importance of comfort and occupancy models for evaluating NV performance, few research articles have considered post-COVID occupancy schedules and adaptive comfort models in NV simulation. To overcome this limitation, this paper integrates the following additional models with the NV simulation: 1) an updated occupancy schedule to reflect post-COVID behaviors in residential buildings and 2) alternative thermal comfort models to account for occupants' thermal sensations inside naturally ventilated or multi-mode residential buildings. 

This research encompasses four sets of experiments conducted using the developed workflow. Firstly, a large population of buildings was generated, and simulations were performed on this population to test the capabilities of the workflow. Secondly, the study investigated the impact of different cooling strategies on energy consumption and occupant thermal comfort across various climate zones. Thirdly, the research explored the influence of building orientation on energy usage. Lastly, the fourth experiment involved visually comparing the hourly changes in energy consumption and occupants' thermal comfort among different cooling strategies to gain insight into their relative performance.

\section{Methodology}
\label{Sec: Methodology}
The workflow developed for this research consists of five steps:

\emph{1- Selecting the modeling parameters}: This involves determining the input parameters that are relevant to the simulation model.

\emph{2- Generation of building geometries}: This step involves creating the 3D model of the building, which is necessary for the simulation process.

\emph{3- Generating the simulation-ready models}: In this step, the building model is converted into a simulation-ready format. The user can choose their desired occupancy and comfort models, or they can use the default values provided by the simulation software.

\emph{4- Conducting the simulation}: The simulation is run using the parameters and model created in the previous steps.

\emph{5- Data analysis on the results obtained}: The results obtained from the simulation are analyzed to draw conclusions and insights.

\begin{figure}[ht!]
     \centering
     \includegraphics[width=3 in]{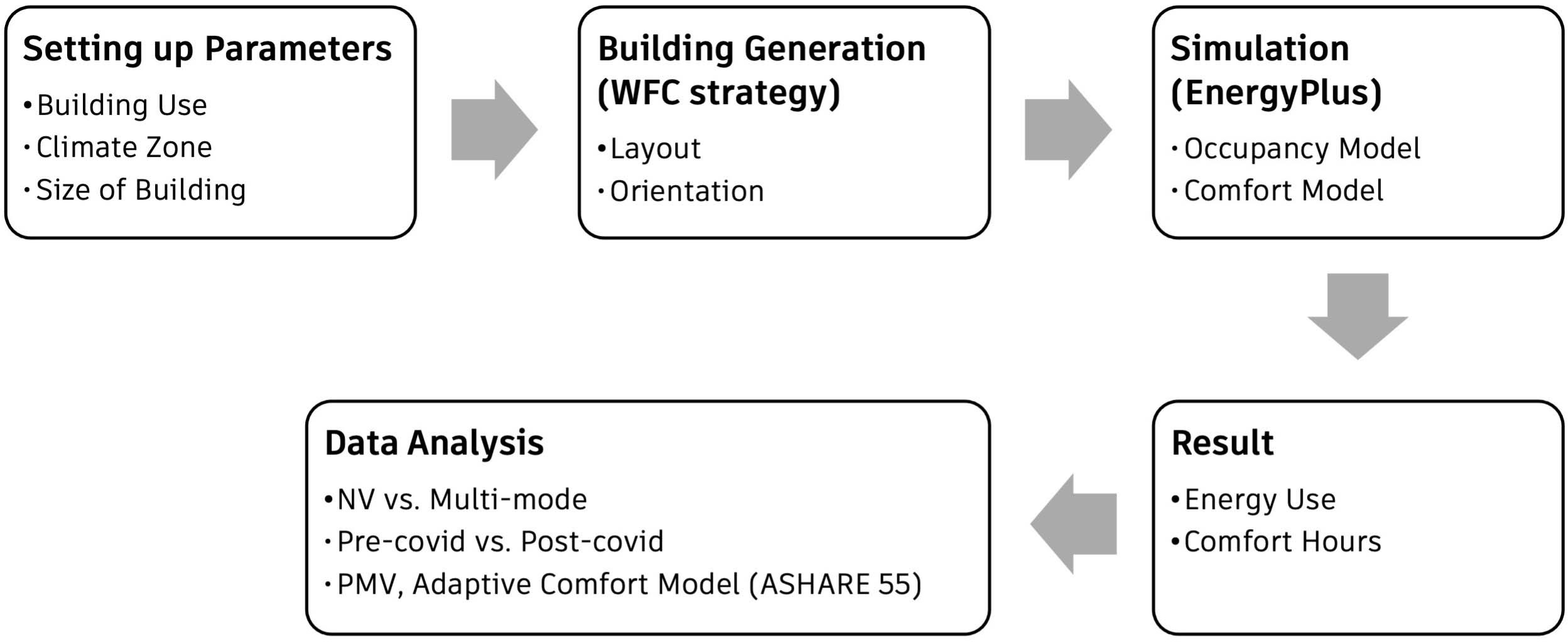}
     \caption{The workflow of the research.}
     \label{method}
 \end{figure}

Figure~\ref{method} illustrates these steps. Each step of the workflow will be discussed in detail in the sub-sections of Section~\ref{Sec: Methodology}.

\subsection{Selecting the modeling parameters}

Firstly, this project set up input parameters for energy and thermal comfort simulation focusing on mid-rise apartments. Over the last three decades, the share of new mid-rise apartment buildings, which have five to twelve floors, has upsurged from 6\% in the 1990s to 41\% of buildings under construction due to stronger demand for urban living and more expensive cost of land than before \cite{sarac2019high}. Considering this trend in the residential building sector, the input parameters for simulating mid-rise apartments are determined based on a technical report written by the National Renewable Energy Laboratory, such as the number of people per area, infiltration rate, energy consumption from lighting per area, etc. \cite{deru2011us}

\begin{figure}[h!]
     \centering
     \includegraphics[width=3 in]{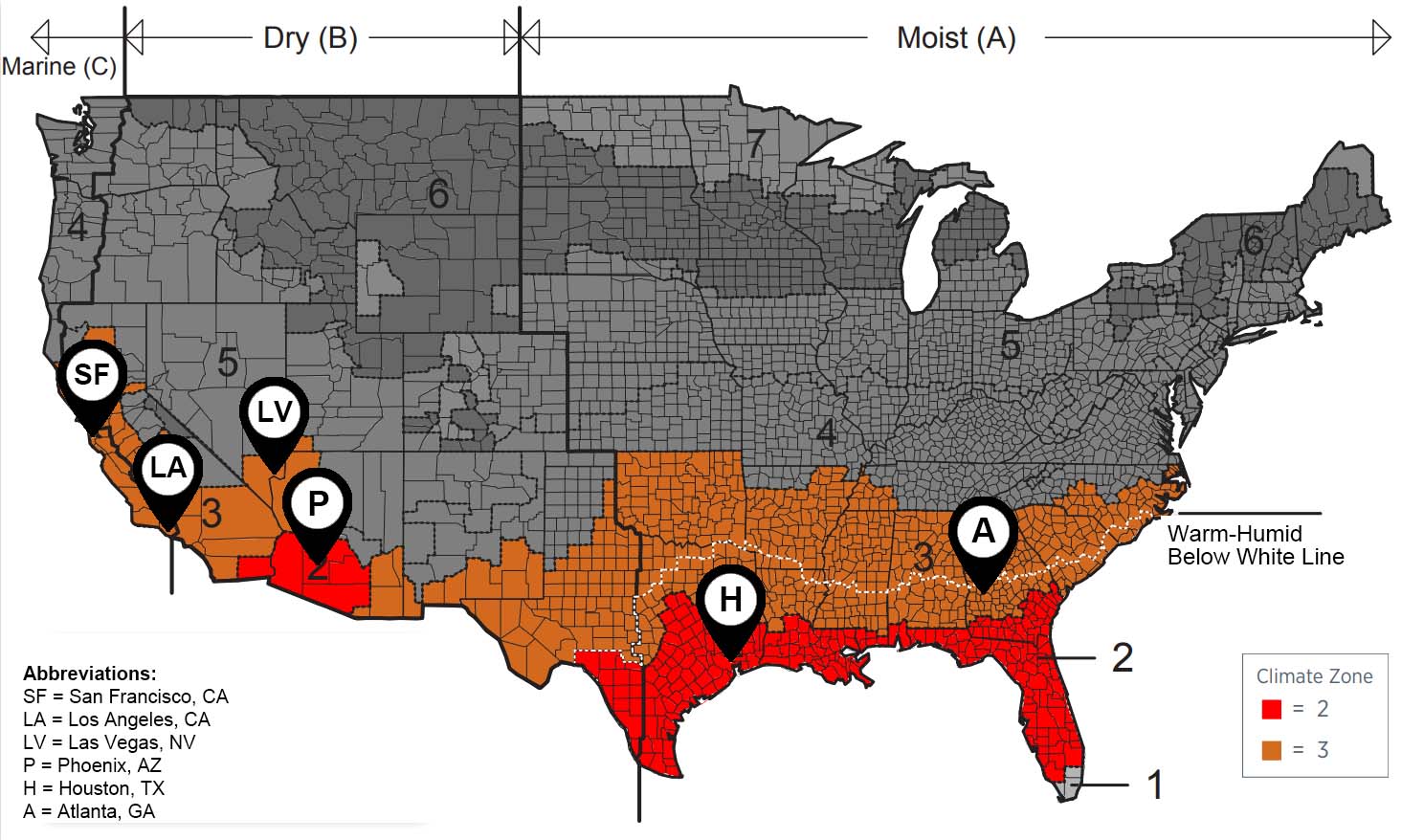}
     \caption{Selection of climate zones}
     \label{climateZones}
 \end{figure}

\begin{figure}[h!]
     \centering
     \includegraphics[width=3 in]{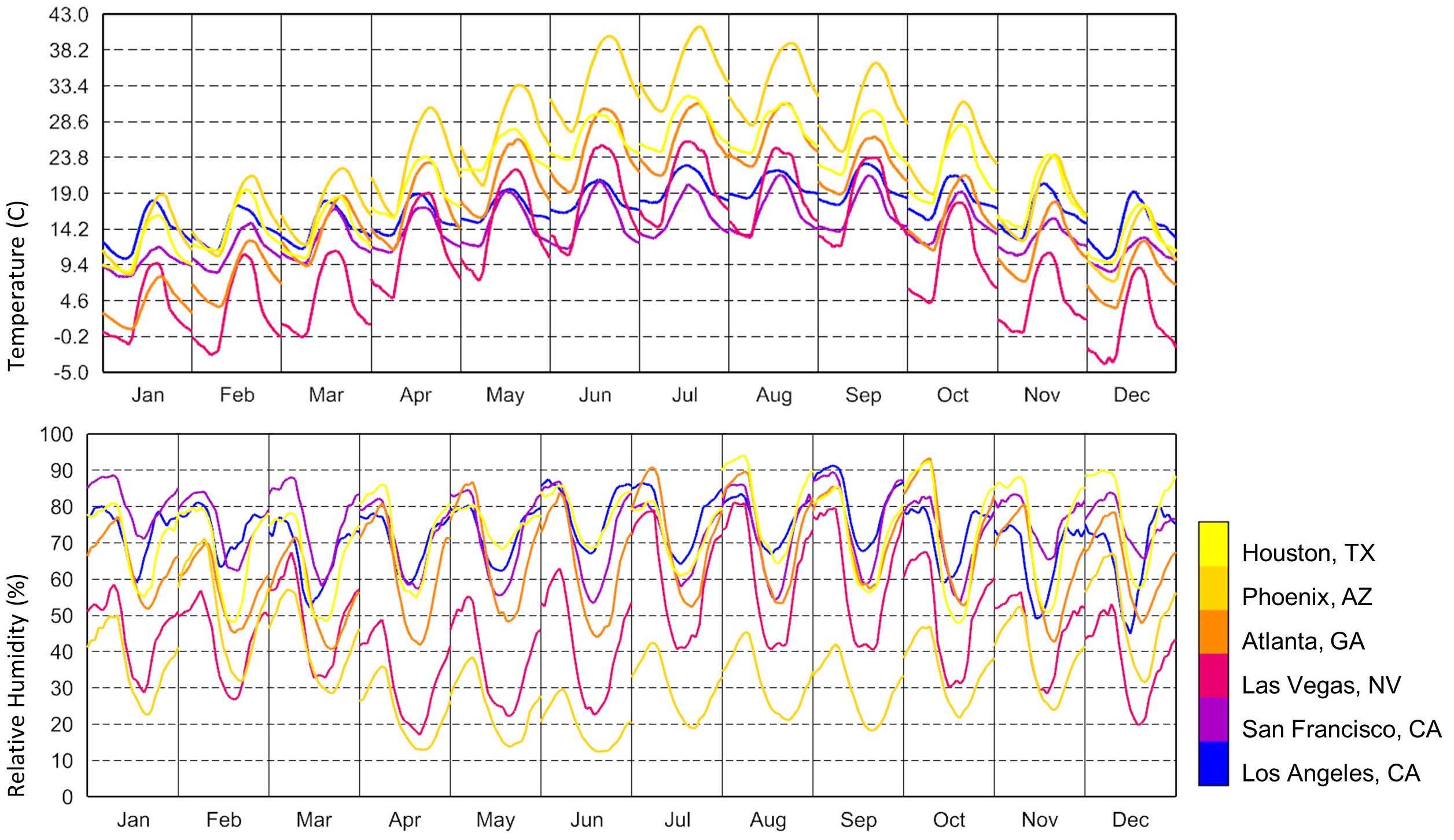}
     \caption{Outdoor temperature and relative humidity of the selected climate zones}
     \label{climateTemperature}
 \end{figure}

Next, climate zones were selected to comparatively analyze the effect of climate zones on a building. Department of Energy and ASHRAE Standard 90.1-2004 classify climate zones into 16 categories based on temperature (Zone 1 to 8) and relative humidity (Zone A to C) \cite{deru2011us,baechler2010building}. Among them, this project mainly focuses on the following six cities to leverage relatively comfortable climates for enhancing indoor thermal comfort, as shown in Figure~\ref{climateZones}: Houston (2A), Phoenix (2B), Atlanta (3A), Los Angeles (3B-CA), Las Vegas (3B-other), and San Francisco (3C). As shown in Figure~\ref{climateTemperature}, each climate shows different patterns of outdoor temperature and relative humidity.
 
\subsection{Generation of building geometries}
\begin{figure}[h!]
     \centering
     \includegraphics[width=3.1 in]{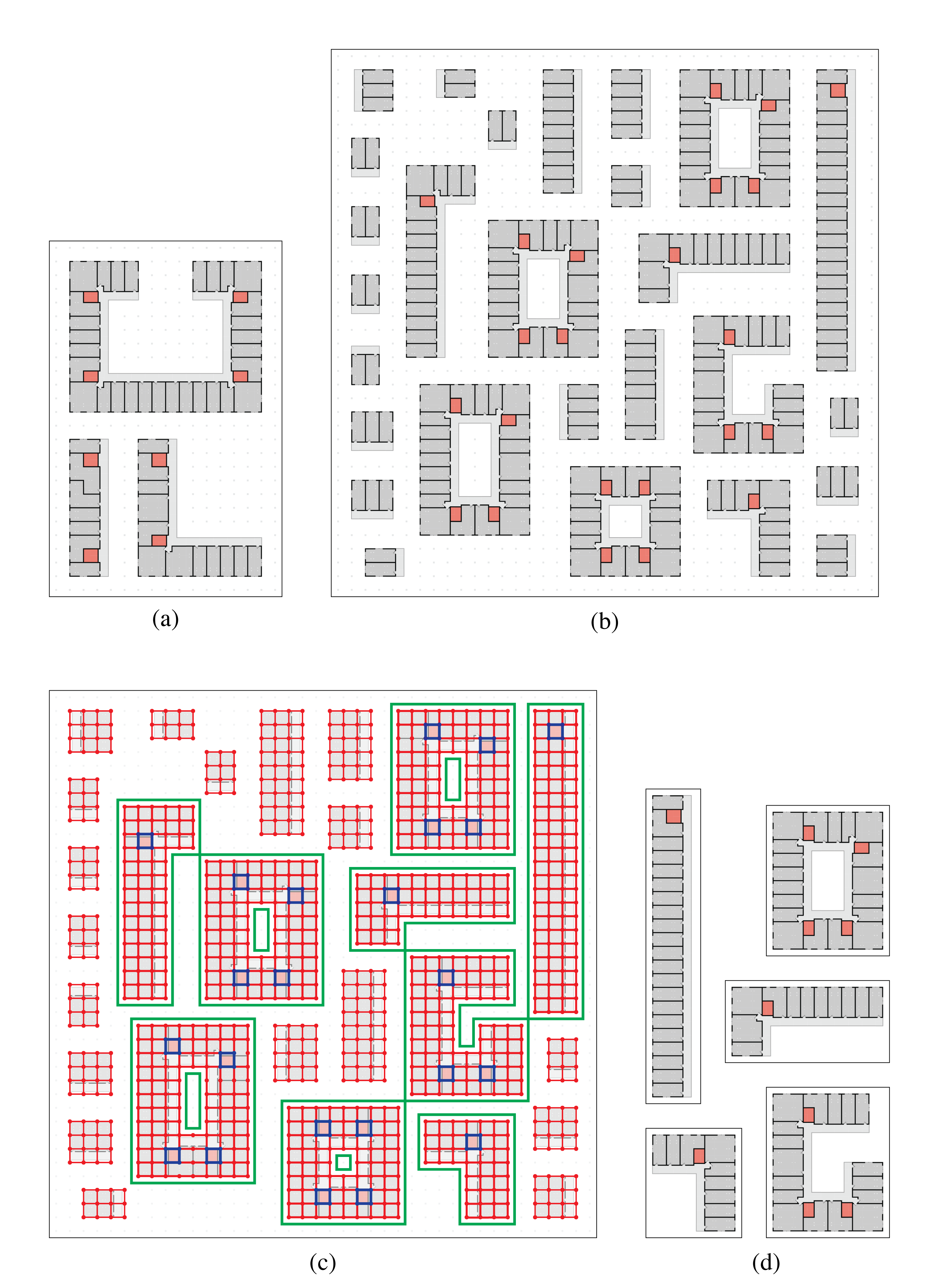}
     \caption{(a) Input example for training WFC model, indicating reference building configurations; (b) Output  40x40 solution from WFC solver containing multiple building layout candidates; (c) Tile grouping using graph analysis with red edges indicate building tile connections, blue edges indicate building core connections, and green outlines indicate valid layouts (containing at least one building core) for extraction; (d) Examples of isolated and extracted building layouts.}
     \label{WFCprocess}
 \end{figure}

To automatically generate building geometries, this project adopts the Wave Function Collapse (WFC) algorithm that enables a geometry-generation model to synthesize constrained design variations using a sparse set of input examples in a manner compatible with multi-objective search methods through variable inputs for tile weights \cite{villaggi2022harnessing,karth2017wavefunctioncollapse,sandhu2019enhancing}. To generate diverse variations of building layouts, the WFC algorithm combines different types of unit tiles—like end wall, side wall, corridor, core, and empty tile—based on rules learned from the provided input samples. First, a series of 300 WFC solutions was generated using a randomized sampling of tile input weights and a 40 x 40 output dimension, each containing multiple candidate building layouts. Second, 553 individual unique building layouts were automatically extracted from the generated solutions using graph-based methods with rotational variants and candidates containing no building cores discarded. Third, feature characteristics were computed for each unique building layout, including counts of building tiles, corners, building cores, building envelope tiles, and individual apartments. Additionally, a series of geometric descriptors were computed including the minimal bounding box area, the normalized aspect ratio, and the compactness of the design (a ratio of building tiles to bounding box area). Figure ~\ref{WFCprocess} illustrates the generation process using the WFC.

\begin{figure*}[ht]
     \centering
     \includegraphics[width=\linewidth]{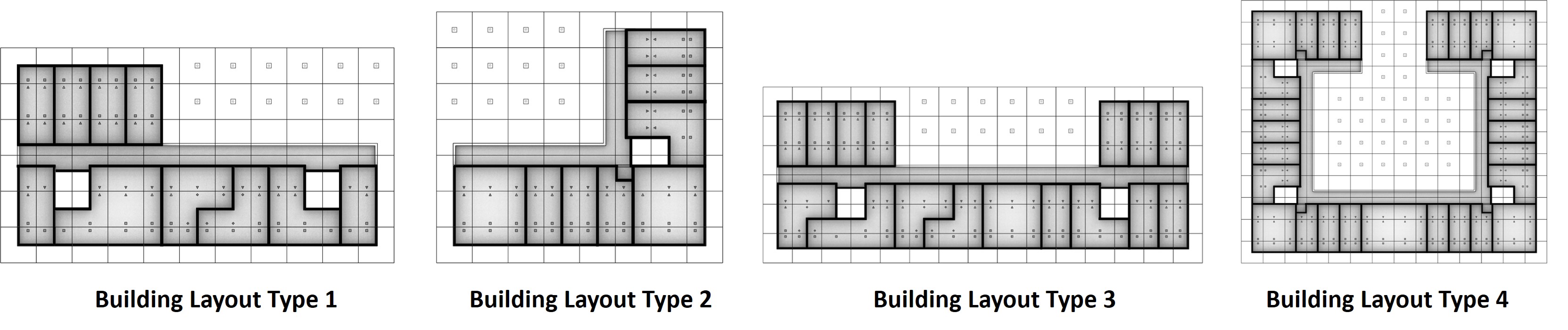}
     \caption{The four building layout types selected for the analysis.}
     \label{buildingLayouts}
 \end{figure*}

From the computed 553 unique candidates, a subset of 189 layouts containing 10 to 15 apartment units was isolated to compare similarly-sized buildings with different layouts to analyze the influential factors on NV. An issue with simulating layouts containing fully-enclosed void space, primarily ring-shaped buildings, led to the exclusion of 13 layouts. After excluding the failed geometries, 176 building layouts are prepared for generating simulation-ready models. As shown in Figure~\ref{buildingLayouts}, we selected four different types of buildings with open corridors for the analysis and comparisons: single-loading corridor, double-loading corridor, L-shaped corridor, and court-yard type. This is because corridors play a significant function in directing outdoor air into the building and facilitating cross ventilation \cite{aflaki2015review,siew2011classification}. 

\subsection{Generating the simulation-ready models}
\begin{figure}[h!]
    \centering
    \includegraphics[width=3.3 in]{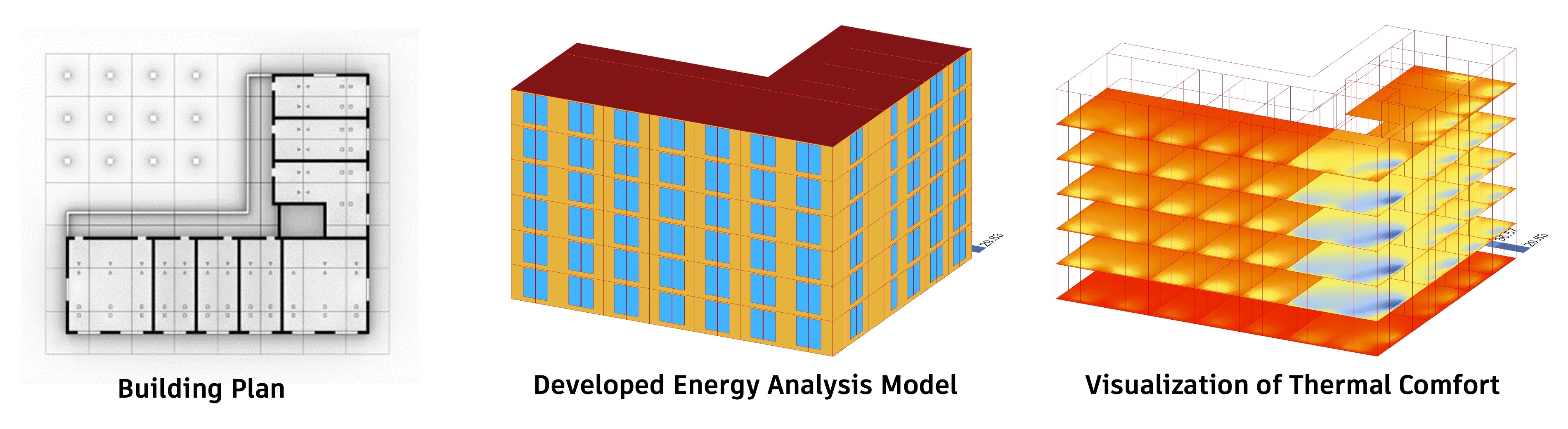}
    \caption{The sample building layout, its developed energy analysis model, and thermal comfort results. }
    \label{buildingSamples}
\end{figure}

As shown in Figure~\ref{buildingSamples}, Grasshopper in Rhinoceros software, Ladybug, and Honeybee plug-ins are used for automating the generation of energy models from the building geometries, and we set up EnergyPlus-based simulation environments combining alternative assumptions for building orientation, heating and cooling strategies, occupancy schedules, and thermal comfort models.

To compare the influence of air-conditioning mode, natural ventilation mode, and multi-mode systems, different parameters are applied: 1) for the air-conditioning mode, all the windows are closed and HVAC system is turned on to ensure the indoor temperature between 21.7°C and 24.4°C, which are setpoints suggested by the NREL technical report \cite{deru2011us}; 2) for the natural ventilation mode, the HVAC system is turned off, and windows are opened or closed to ensure the room temperatures between 21°C and 28°C. Windows are opened or closed only when outdoor temperature is at least 3°C lower than indoor temperature as the main purpose of the window operation is for space cooling. The windows are operated while the outside temperature is in the range of 18°C to 25°C. 3) for the multi-mode, windows are operable same as NV mode. HVAC will be running if the indoor temperature is lower than 20°C or higher than 29°C. We assumed a 1-degree buffer zone between the temperature thresholds of HVAC and NV modes to avoid unnecessarily frequently switching between AC and NV modes around the temperature thresholds.

\begin{figure}[h!]
    \centering
    \includegraphics[width=3 in]{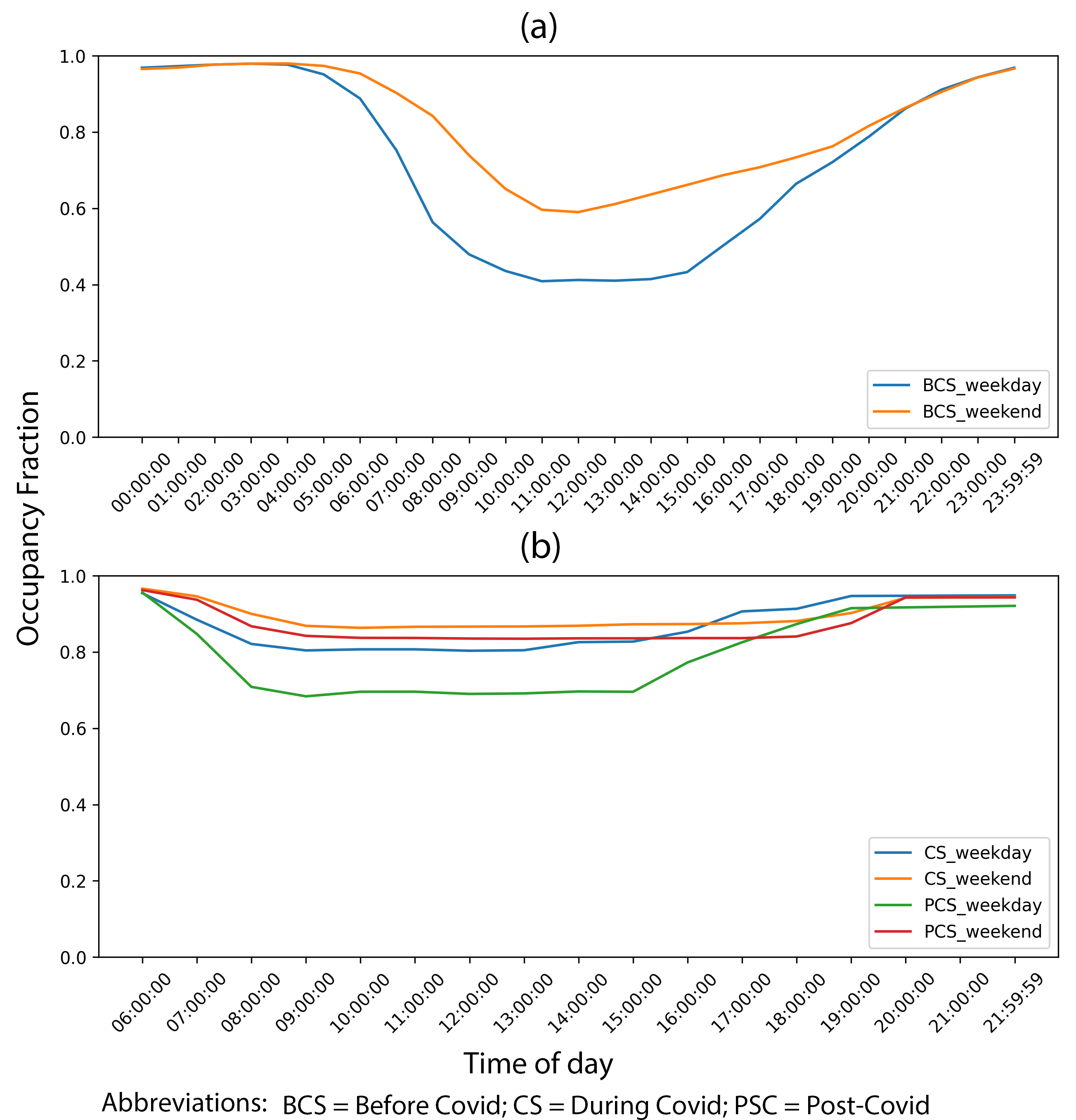}
    \caption{Average occupancy schedule before, during, and after COVID-19 pandemic on weekdays and weekends (regenerated based on the collected data from Mitra et al. (2021))}
    \label{occupancySchedule}
\end{figure}

To apply realistic pre-COVID and post-COVID occupancy patterns, occupancy schedule data was obtained from \citeN{mitra2021variation}. For developing pre-COVID US residential occupancy profiles, the researchers set up their dataset based on the American Time Use Survey and Current Population Survey conducted by the US Census Bureau; for the post-COVID residential occupancy profiles, they implemented field survey data. This dataset provides occupancy fraction information prior to, during, and after the COVID-19 pandemic period, and each group is subdivided into weekday/weekend and high/middle/low-income levels. As a preliminary study, we mainly focus on the occupancy fractions for the three periods with middle-income levels, as shown in Figure~\ref{occupancySchedule}. The occupancy fraction value ranges from 0 to 1, where 0 represents no occupant is at home; 1 represents all occupants present.

For evaluating occupants’ comfort based on the simulated indoor environment, PMV and adaptive comfort models are adopted, and their approaches are different. The PMV model estimates comfort values based on the heat balance model by considering indoor temperature, thermal radiation, relative humidity, air velocity, occupants' metabolic rate, and clothing insulation \cite{international2005ergonomics}. On the other hand, the adaptive comfort model calculates the optimal comfort temperature based on the mean outdoor dry bulb temperature \cite{de2002thermal}. 

\subsection{Conducting the simulation}
The workflow provides simulation results as two metrics: energy use intensity (EUI) and percentage of neutral time (PNT). EUI is the total energy consumption per year divided by the gross floor area of the building. That is, the low EUI of the building represents high energy performance. On the other hand, PNT represents the percentage of time when the comfort level is within acceptable limits that vary depending on comfort models. These two metrics are mainly used for comparatively analyzing the performance of the energy models with different heating/cooling strategies, comfort models, and occupancy models. Please note that the PNT results shown in the paper are based on Adaptive Comfort Model (ACM); this model has shown higher accuracy for predicting thermal comfort in naturally ventilated buildings and better reflects the psychological dimension of occupants' adaptation inside their homes than the PMV model \cite{de2002thermal}.

\section{Results}

\subsection{Simulating a set of buildings with diverse forms and orientations}
This first study involved taking the 176 generated building layouts and simulating their performance for each of the 12 orientation angles. The investigated models included generated layouts that were oriented at angles ranging from 0 through 330 degrees with a step size of 30 degrees.

\begin{figure}
    \centering
    \subcaptionbox{}{\includegraphics[width=0.49\linewidth]{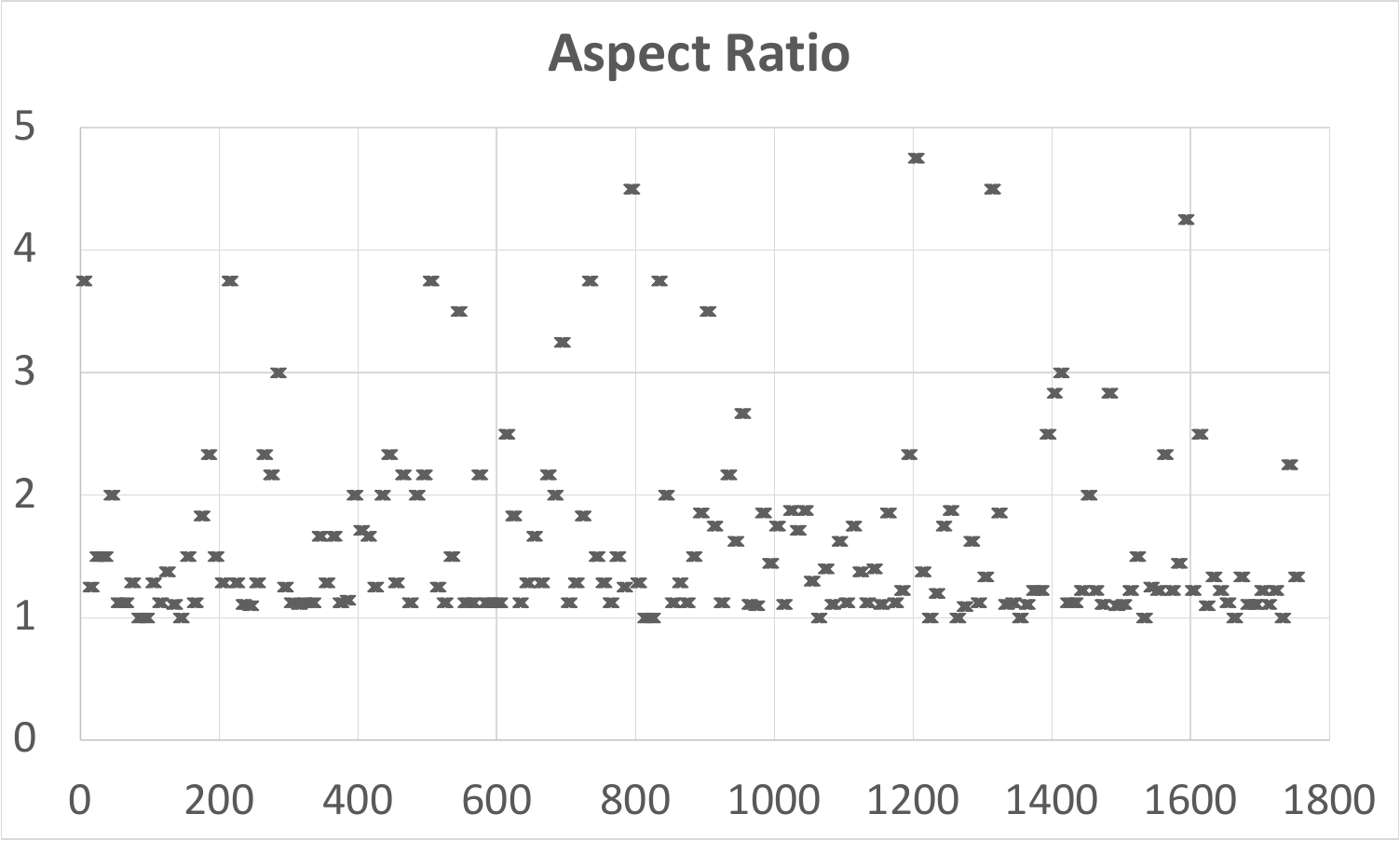}}
    \subcaptionbox{}{\includegraphics[width=0.49\linewidth]{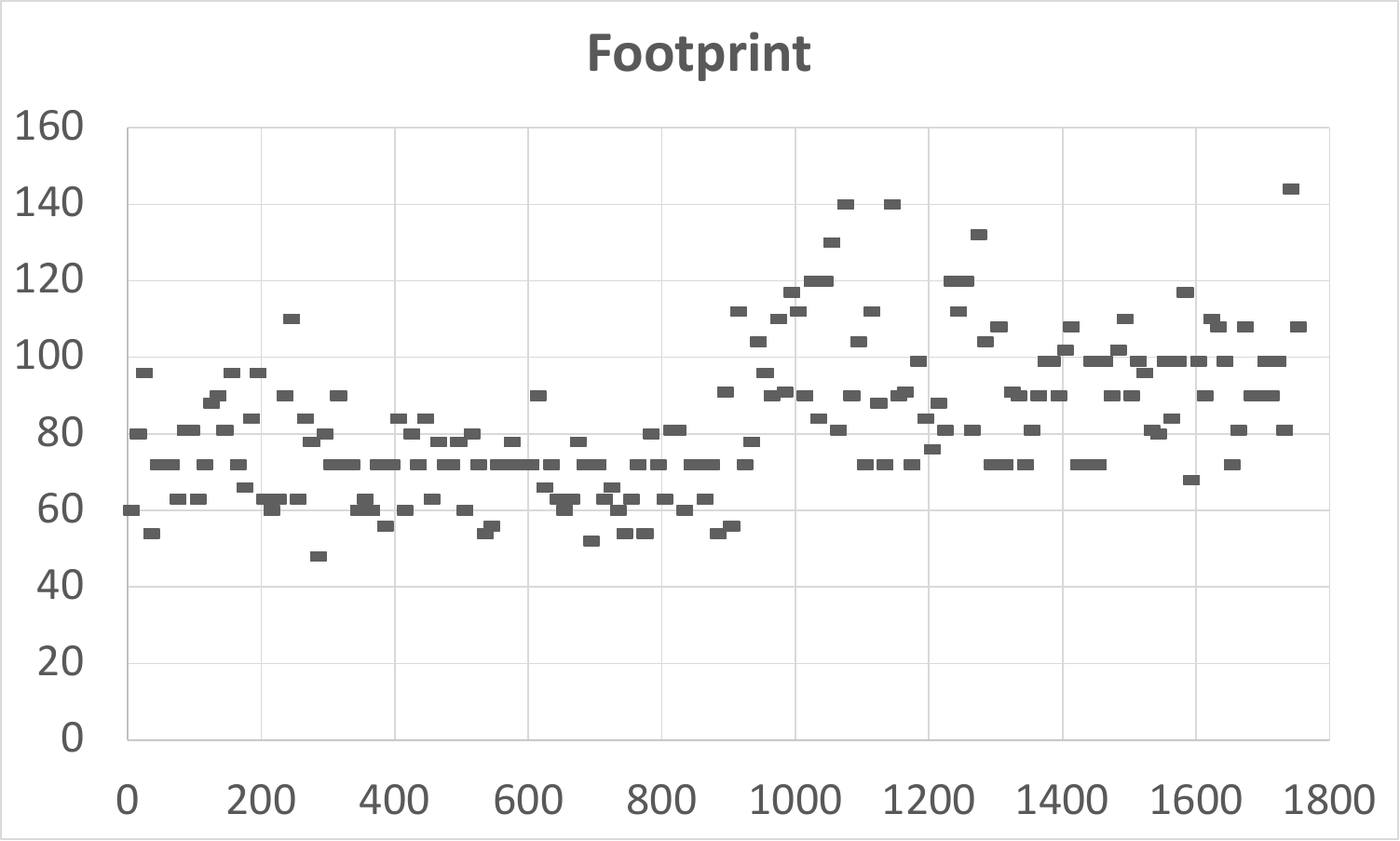}}
    \subcaptionbox{}{\includegraphics[width=0.49\linewidth]{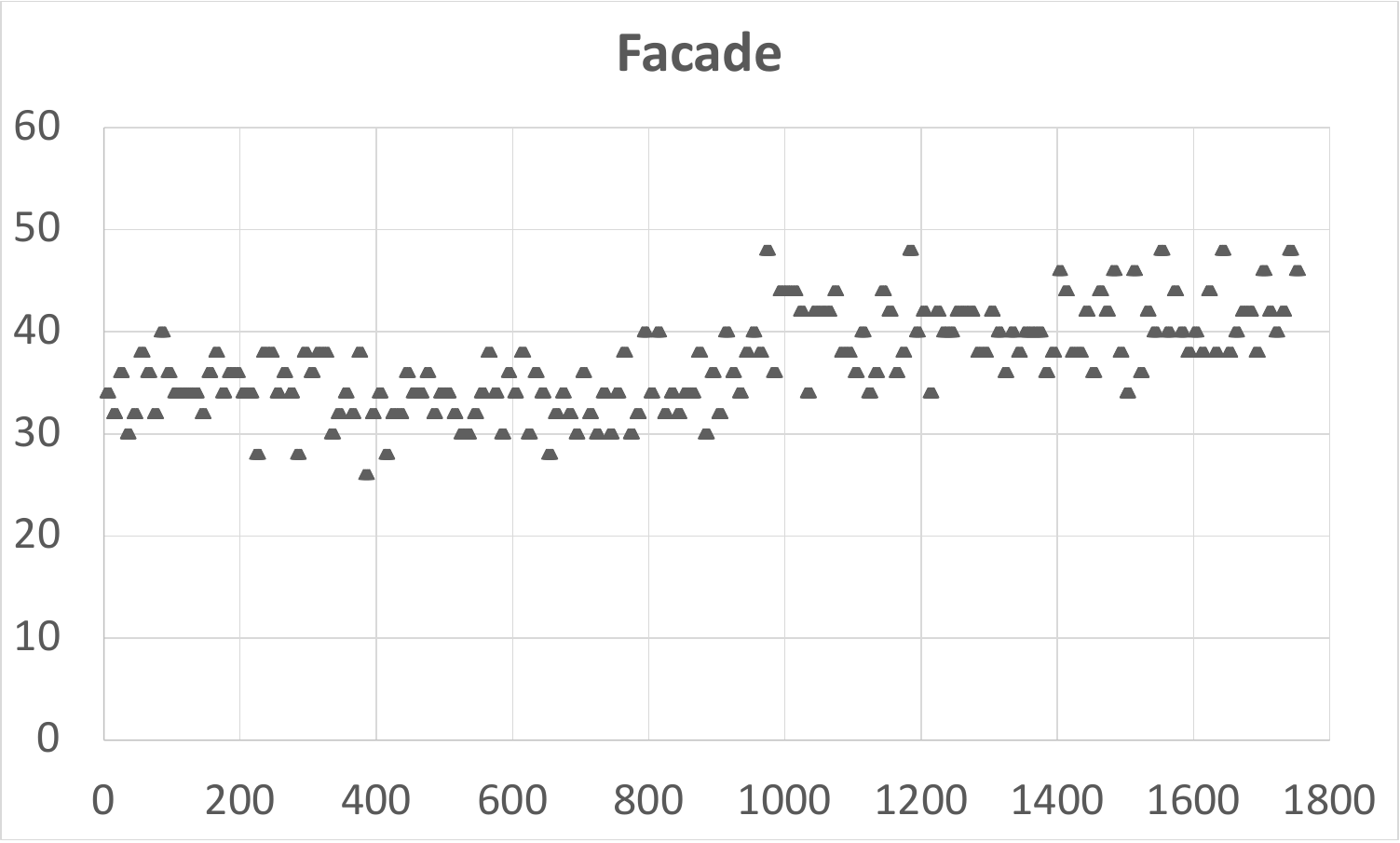}}
    \subcaptionbox{}{\includegraphics[width=0.49\linewidth]{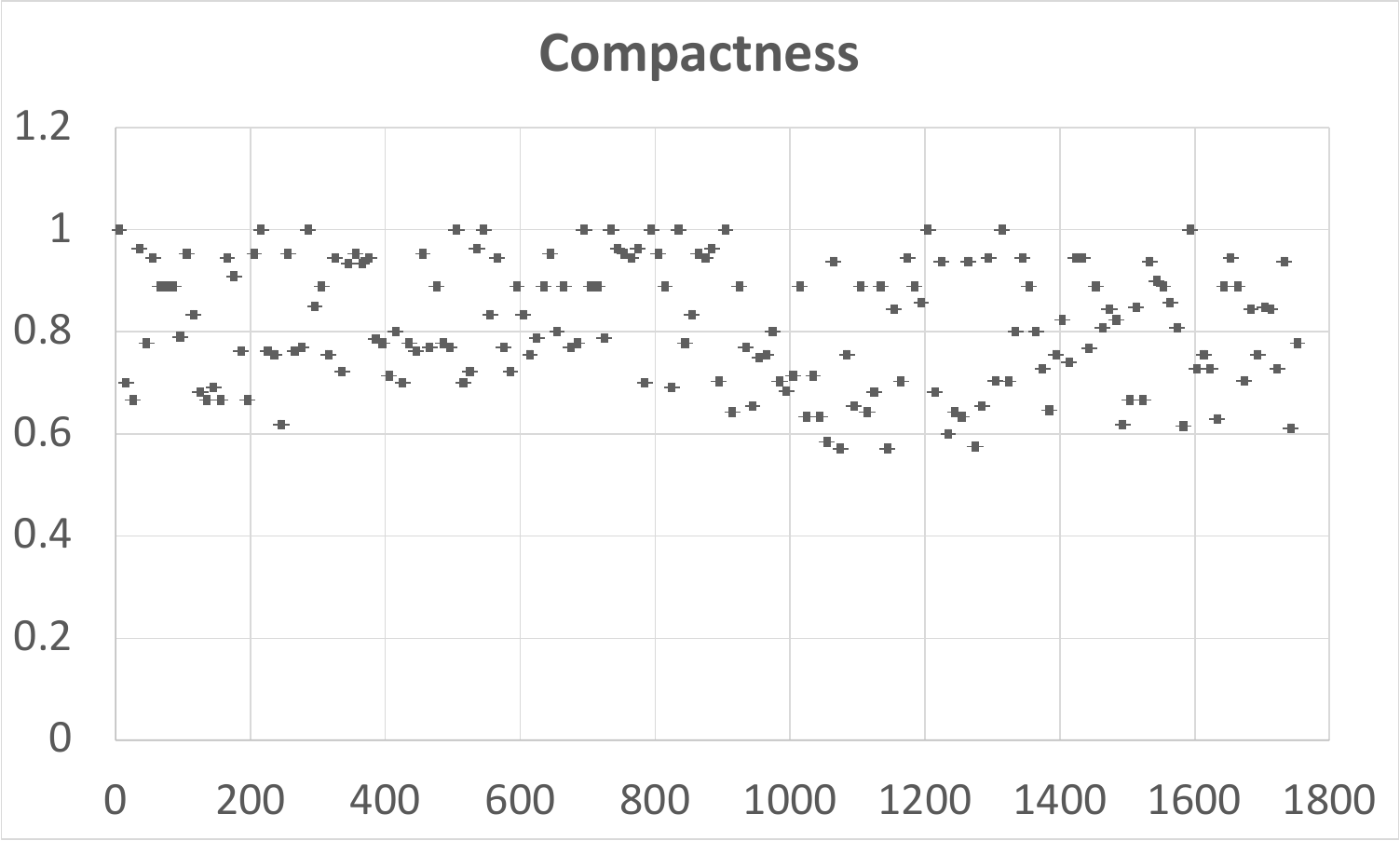}}
    \caption{(a) Aspect Ratio, (b) Footprint, (c) Facade, (d) Compactness values for all generated and modeled buildings. The x-axis of the plots represents the ID-numbers of the buildings in the population.}
    \label{buildingsdDiversity}
\end{figure}

Figure~\ref{buildingsdDiversity} illustrates the Aspect Ratio, Footprint, Facade, and Compactness measures of the generated building population; the x-axis of the images shows the ID-number of the buildings in this population. In this context, Aspect Ratio refers to the normalized aspect ratio for the building footprint, calculated as the maximum of the width and length divided by the minimum. Footprint refers to the minimal bounding box area of the building pattern (width multiplied by length including open space). Facade refers to the count of tiles corresponding to the building envelope, i.e., the outer perimeter of the buildings. Compactness refers to the percentage of the footprint occupied by building tiles, calculated as the count of tiles corresponding to building elements divided by the footprint.

The simulations were run for all the buildings using the MM cooling strategy. The simulations were conducted for the Phoenix, AZ location, which is classified as climate zone 2B (hot and dry). The occupancy presence model considered a post-COVID scenario and households with a middle-income level.

\begin{figure*}[ht]
    \centering
    \includegraphics[width=\linewidth]{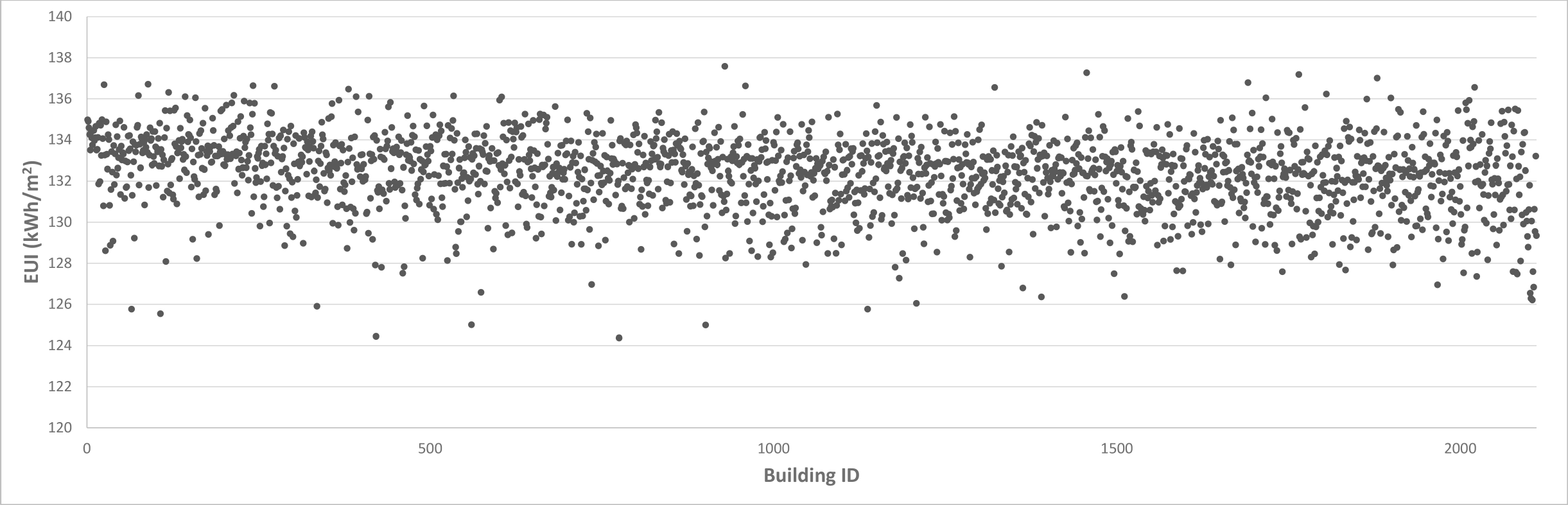}
    \caption{The EUI obtained for all of the simulated buildings, assuming: Phoenix, AZ location, MM cooling strategy, a post-COVID occupancy presence model, and middle-income households. }
    \label{EUIall}
\end{figure*}

The results obtained for EUI are displayed in Figure~\ref{EUIall}, showing a 10\% difference in EUI within the dataset. The study also observed a 16\% and 22\% difference in the Predicted Mean Vote (PMV) and Adaptive Comfort model, respectively, for the buildings' PNT, obtained based on Adaptive Comfort Model. The results obtained for PNT are not shown here.

\subsection{Exploring 3 cooling strategies in 6 climate zones}
\begin{figure}
    \centering
    \includegraphics[width=3.1 in]{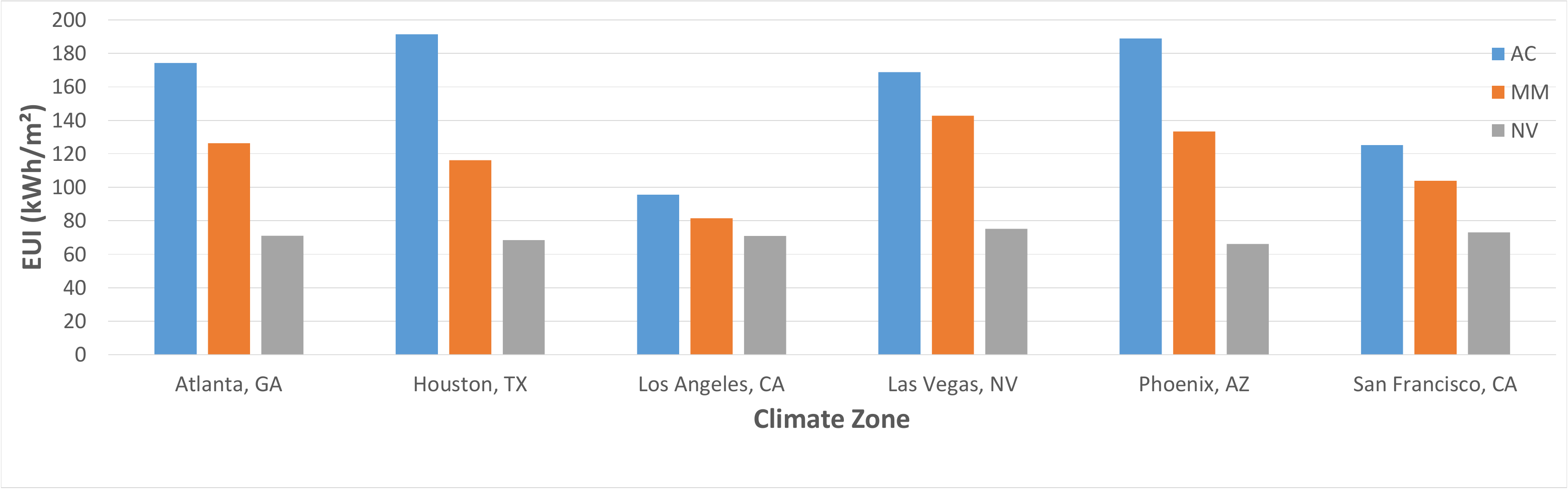}
    \caption{The EUI values obtained for the 6 climate zones, 3 cooling strategies, and Type-2 building layout. A lower EUI indicates higher energy efficiency and a more sustainable building design.}
    \label{Fig: EUI-l-shape-6climate}
\end{figure}

\begin{figure}[h!]
    \centering
    \includegraphics[width=3.1 in]{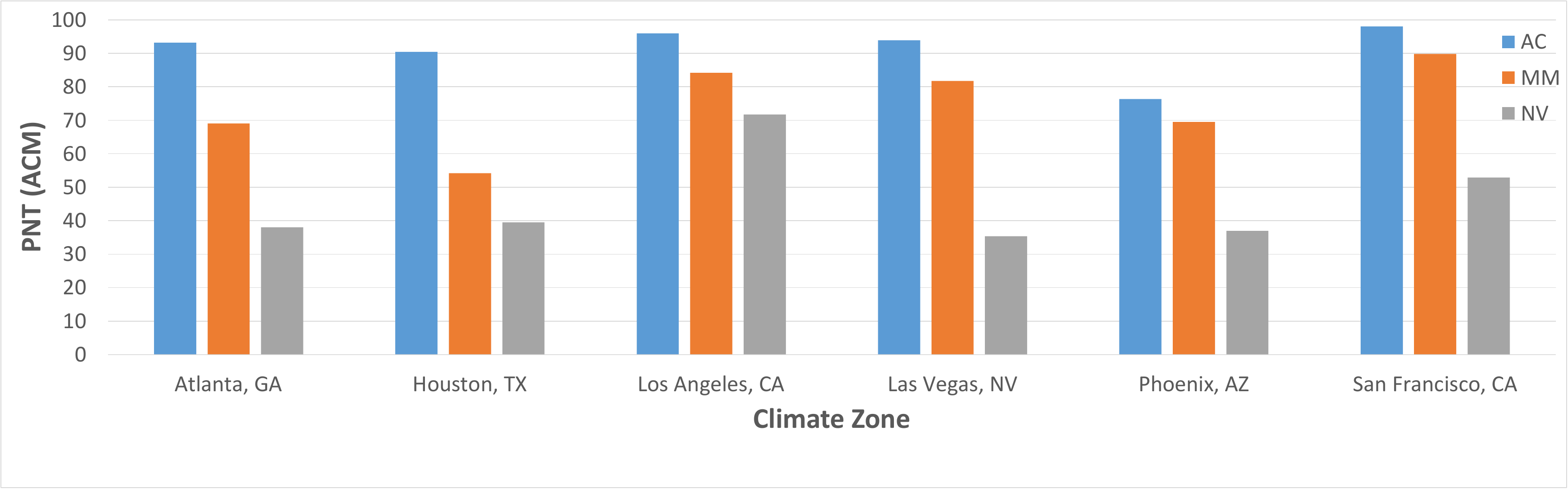}
    \caption{The PNT obtained for the 6 climate zones, 3 cooling strategies, and a Type-2 Buildings Layout. A higher PNT indicates a greater level of thermal comfort and suggests that the building design is successful in meeting the occupants' comfort needs.}
    \label{Fig: PNT-l-shape-6climate}
\end{figure}

Figures~\ref{Fig: EUI-l-shape-6climate} and \ref{Fig: PNT-l-shape-6climate} provide a comparison of the EUI and PNT data obtained for 6 different climate zones. These simulations were conducted using the assumption of a post-COVID occupancy presence model and households with a medium income level. The simulations were conducted on Building Layout Type 2 shown in Figure~\ref{buildingLayouts}.

Across all climate zones, it was observed that the NV and MM modes resulted in lower energy consumption compared to the HVAC mode. However, the use of NV and MM modes led to a decrease in PNT. Although this reduction may not be ideal, it is important to note that the acceptability of the PNT level depends on the specific region and its comfort standards.

\subsection{Exploring the impact of building orientation on energy usage}
\begin{figure}[h!]
    \centering
    \includegraphics[width=3.1 in]{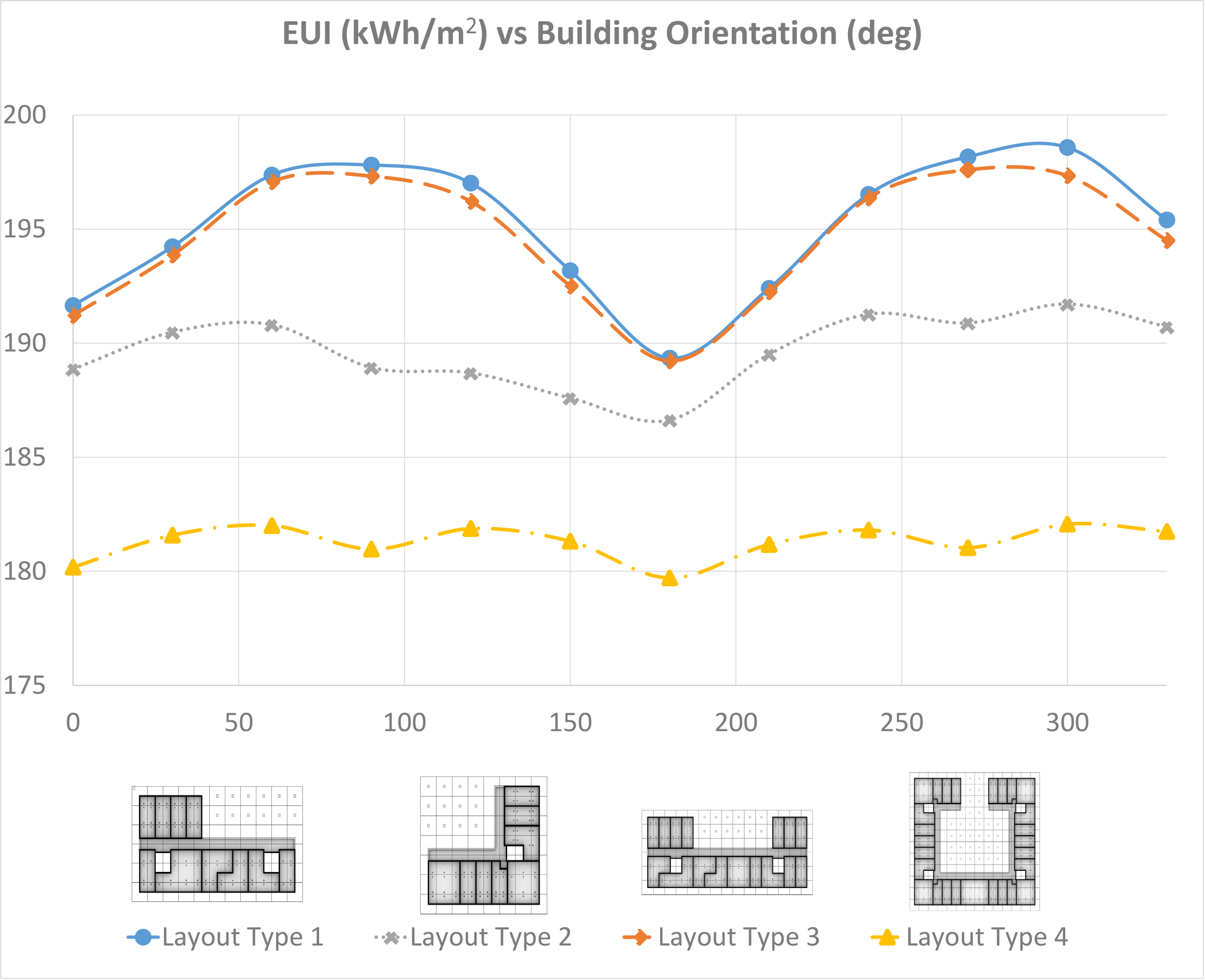}
    \caption{Impact of building form and orientation on EUI. The trend and the level of EUI change depends on the orientation as well as building form. }
    \label{EUI-orien-Type}
\end{figure}

The study examined the impact of building orientation on energy usage for the 4 building layouts presented in Figure~\ref{buildingLayouts}. Figure~\ref{EUI-orien-Type} displays the EUI obtained for these buildings when oriented at angles ranging from 0 through 330 degrees with a step size of 30 degrees. These simulations were conducted for the Phoenix, AZ location and using an AC cooling mode.

The results highlight that both the trend and level of EUI change depending on the orientation and building form. This suggests that while general rules of thumb can provide rough estimates on energy savings through building orientation, considering the specific building form can provide better estimations, even during the early stages of design. 

\begin{figure}[h!]
    \centering
    \includegraphics[width=3.1in]{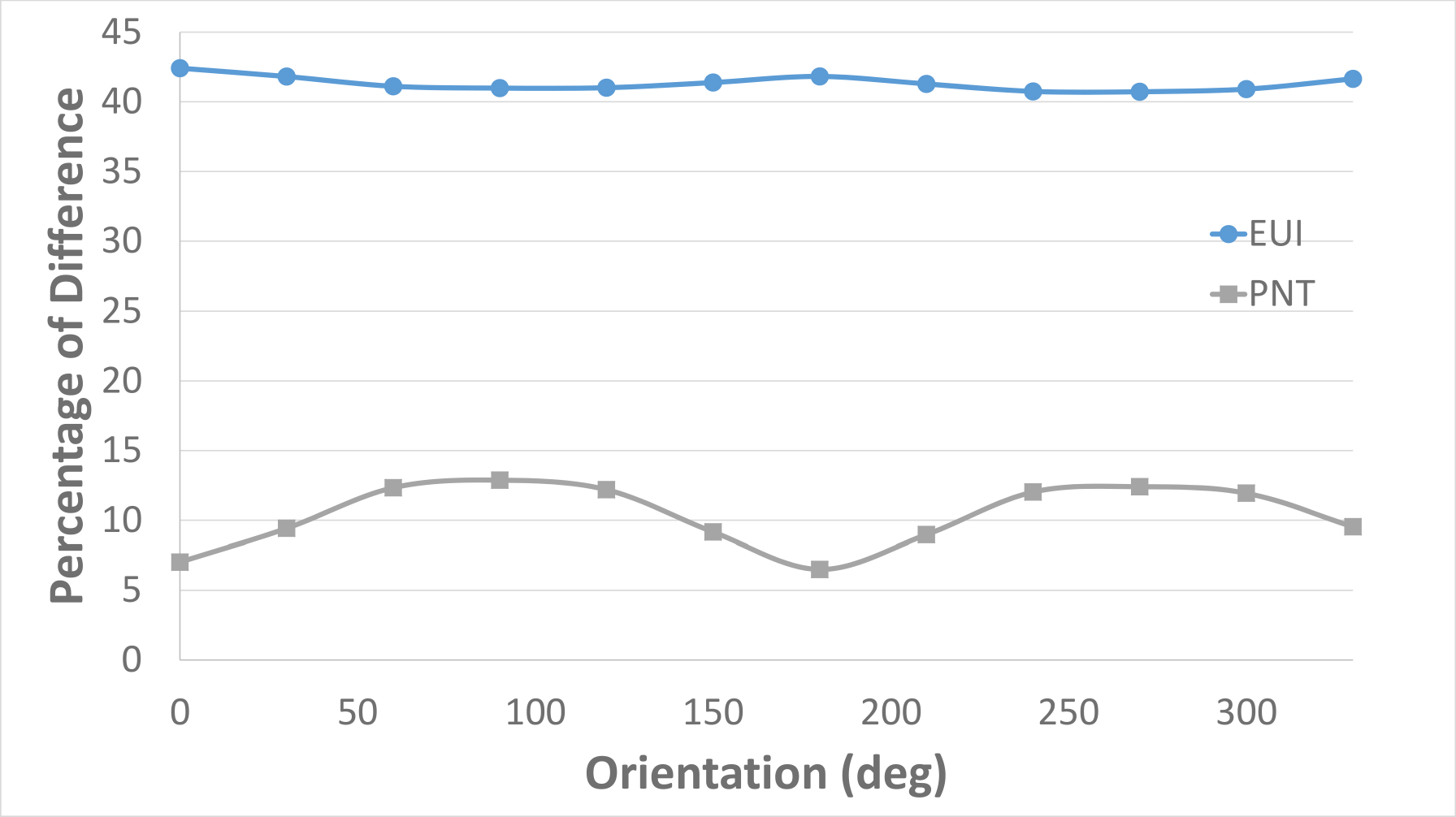}
    \caption{The percentage difference in the EUI and PNT levels obtained for a Type 3 Building Layout, comparing the AC mode and MM mode at different orientations. }
    \label{EUI-PNT-Perc-Diff}
\end{figure}

The percentage difference in the EUI and PNT levels obtained for Type 3 Building Layout, presented in Figure~\ref{buildingLayouts}, comparing the AC mode and MM mode at different orientations, is presented in Figure~\ref{EUI-PNT-Perc-Diff}. The simulations were conducted for the Phoenix, AZ location.

The results show a significant reduction in EUI, approximately 40\%, when the MM mode is utilized. On the other hand, the decrease in PNT is only 10\%. It is important to note that the change in PNT level depends on the selected thermal comfort model. In this case, an Adaptive Thermal Comfort model was employed for the comparison and analysis.

\subsection{Exploring hourly EUI and PNT for 3 cooling strategies}
\begin{figure*}[ht]
    \centering
    \includegraphics[width=4.4in]{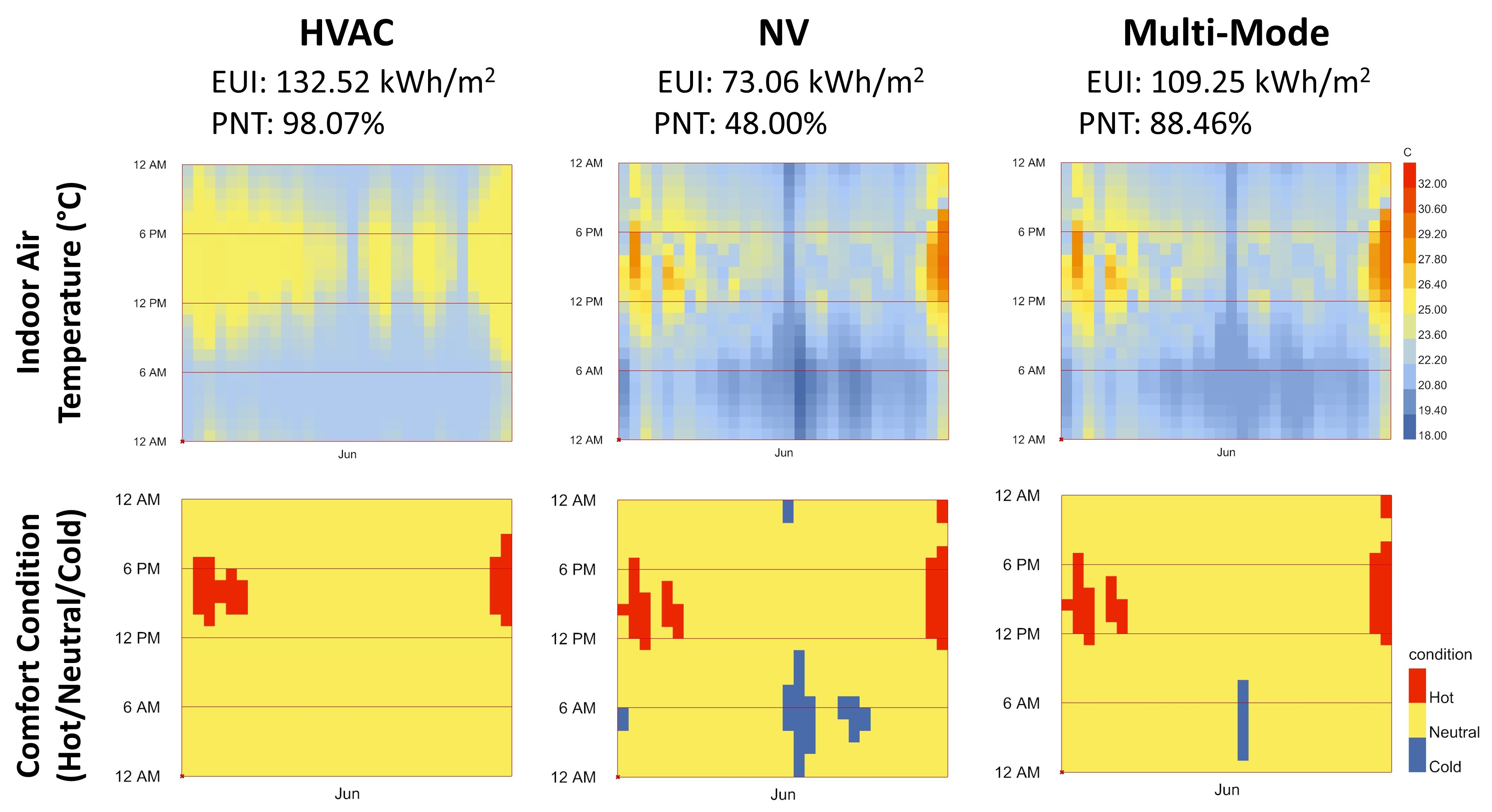}
    \caption{A visual comparison of the hourly values of EUI and PNT for the month of June in a simulation case for Layout Type 2, assuming a San Francisco location}
    \label{visualization-results-EUI-PNT}
\end{figure*}

Figure~\ref{visualization-results-EUI-PNT} provides a visual comparison of the hourly values of EUI and PNT for the month of June in a simulation case for Building Layout Type 2, assuming a San Francisco location. The results show that the MM cooling strategy reduces EUI by approximately 11\% compared to the HVAC strategy, while still maintaining reasonable PNT values above 85\%.

Preliminary results for Houston and Atlanta (not shown in the figure) indicate even larger potential decreases in energy consumption levels: 32.5\% for Houston and 22.1\% for Atlanta. However, the simulations suggest that natural ventilation alone is not sufficient to maintain the indoor climate within an acceptable range.

The images in Figure~\ref{visualization-results-EUI-PNT} illustrate the hours of the day with poor PNT levels, indicating either hot or cold sensations for the occupants, as well as the times with peak energy consumption levels. This information can be utilized in the design of smart controllers and advanced energy-saving strategies and technologies. For instance, green energies can be generated/collected and stored during specific hours of the day and used to power the MM system during hours with low PNT, optimizing both energy usage and occupant comfort.

\section{Discussion}
This research presents a developed workflow for generating, modeling, and simulating diverse building types and layouts with different orientations, specifically focusing on incorporating passive cooling techniques, such as NV, in the early design stages of sustainable household apartment buildings. The workflow consists of five steps: selecting modeling parameters, generating building geometries using the WFC algorithm, converting the models into a simulation-ready format, conducting the simulations, and analyzing the obtained results. The workflow offers a range of simulation options, including different ventilation methods (NV, MM, AC), occupancy presence models, and thermal comfort assessment models. The simulation results are evaluated in terms of the EUI, representing energy consumption per square foot per year, and the PNT, representing the percentage of time when occupants experience a comfortable indoor environment.

The research experiments in this project aimed to provide insights into the impact of utilizing different cooling strategies and support the design of more energy-efficient and comfortable residential buildings. The experiments assessed the impact of cooling strategies on energy consumption and thermal comfort by comparing air-conditioning, natural ventilation, and multi-mode cooling strategies, considering various parameters. To address the limitations of existing studies in evaluating NV performance, the research integrates updated occupancy schedules to reflect post-COVID behaviors and alternative thermal comfort models for assessing occupants' thermal sensation in naturally ventilated or multi-mode residential buildings. 

The first research experiment generated and simulated a large set of buildings. The results highlighted that the large pool of data generated by the simulations can be challenging for humans to digest. While optimization engines can assist in the process of optimal design selection or knowledge inference, they are probably not sufficient on their own. Therefore,  AI-driven technology has the potential to greatly benefit designers and stakeholders in tailoring specific buildings by rapidly exploring different options for larger design projects and surfacing near-optimal designs. The technology can also be utilized to extract and present knowledge from the data in various interactive formats, including text, images, and video, to aid decision-making processes. Moreover, a few high-performing building design options can be selected with and passed on to the next design stages. Further detailed analysis can be conducted on these cases by providing additional design information available in later design stages. The selected cases can also be moved to high-fidelity analysis tools such as computational fluid dynamics (CFD) for more detailed analysis.

The second research experiment indicated that both the NV and MM cooling strategies result in lower energy consumption compared to the conventional HVAC mode across all of the examined climate zones. On the hand, the use of NV and MM modes lead to a decrease in PNT, and it is important to observe that NV alone may not adequately maintain the indoor climate within an acceptable range. Our results suggest that implementing the MM mode could lead to substantial improvements in energy efficiency, with a noteworthy decrease in EUI of about 18\% to 40\%. Although there is a slight reduction in thermal comfort (PNT-ACM) compared to the AC mode, it is important to consider the overall trade-off between energy efficiency and occupant comfort. It is worth noting that the acceptability of the reduced PNT level depends on the specific region and its comfort standards. The contribution of MM mode to energy saving is aligned with previous research works. \citeN{gonzalez2016energy} reported 13\% saving of total cooling energy combining HVAC with NV in social housing in Spain, and \citeN{weerasuriya2019holistic} and \citeN{yik2010energy} showed 8-27\% and 25-50\% reduction of the energy consumption of residential buildings in Hong Kong, respectively. However, the proposed workflow has room to increase energy savings by optimizing window size, window position, thermal insulation, etc. Additional design variables will be included for improving energy efficiency in future work. 

The third experiment shows that considering the orientation and form of a building is highly advisable for accurate estimations of energy savings during the early design stages. While general rules of thumb provide rough estimates, the research demonstrates that the trend and level of EUI vary based on these factors. It is important to consider the interaction between building form and orientation when aiming for energy efficiency. Simply relying on general guidelines for building orientation may not be sufficient, as the specific form of the building can have a notable influence on the EUI. Therefore, a comprehensive approach that takes into account both building form and orientation is necessary to achieve accurate estimations and optimize energy performance during the early design process.

The fourth research experiment highlights hourly simulation results and determines the hours of the day with poor PNT levels, indicating discomfort for occupants, as well as the peak energy consumption times. This information can be utilized to design smart controllers and advanced energy-saving strategies, such as generating and storing green energy during specific hours to power the MM system when the PNT is low. Such an optimization approach would aim to effectively balance energy usage and occupant comfort over time.

It should be noted that the results obtained in this work are approximate and may vary depending on specific building characteristics and simulation conditions. The goal of this work was to provide reasonable evaluations for the energy consumption and occupants' thermal comfort inside buildings at early design stages where limited design data is available.

In the future, we are planning to expand the capabilities of the developed workflow by providing more simulation options including controllers for running the MM and AC systems. In addition, we aim to develop a user interface for the workflow. On the experimental side, we are planning to develop a surrogate model based on the simulation results to let designers and decision-makers infer knowledge from the pool of the data in a more efficient way. The model could be also incorporated into a tool to speed the search for optimal sustainable designs.

\section{Conclusion}
This research introduces a developed workflow for generating, modeling, and simulating diverse building types and layouts with different orientations, with a focus on incorporating passive cooling techniques like natural ventilation (NV) in the early design stages of sustainable household apartment buildings. The workflow was used to perform four research experiments in order to evaluate the impact of cooling strategies on energy consumption and thermal comfort of occupants inside buildings. 

The findings emphasize the benefits of implementing MM cooling strategy, as it significantly improves energy efficiency with a notable decrease in EUI. Although there is a slight reduction in thermal comfort compared to the AC mode, the trade-off between energy efficiency and occupant comfort must be considered. However, the study highlights that relying solely on NV may not be sufficient to maintain an acceptable indoor climate. The findings demonstrate the importance of considering the interaction between building form and orientation when aiming for energy efficiency. Simply relying on general guidelines for building orientation may not be sufficient, as the specific form of the building can have a notable influence on the EUI.

While the use of the WFC algorithm can help generate a diversity of design options that vary in their interior layouts as well as orientation, the large amount of data produced by the ensuing simulations may be challenging for architects to interpret. To aid the decision-making process, knowledge extraction from the data and AI-driven technology could be highly beneficial for designers and other stakeholders. These technologies can provide information in interactive formats facilitating tailored building design and exploration of options for larger-scale projects.

\bibliographystyle{achicago}
\bibliography{xbib}

\end{document}